\documentclass[a4paper,12pt]{article}
\pdfoutput=1 

\usepackage{a4wide}
\usepackage{cite}

\usepackage[T1]{fontenc}
\usepackage[latin1]{inputenc}

\usepackage{empheq}
\usepackage{amssymb}
\usepackage{amsmath}
\usepackage{amsfonts}
\usepackage{xcolor}

\DeclareMathAlphabet{\mathpzc}{OT1}{pzc}{m}{it}

\numberwithin{equation}{section}

\usepackage[colorlinks]{hyperref}

\hypersetup{
 citecolor=blue,
 linkcolor=blue,
 urlcolor=blue}

\newcommand{\lp}{\left(}
\newcommand{\rp}{\right)}

\newcommand{\ep}{\epsilon}
\newcommand{\be}{\begin{equation}}
\newcommand{\ee}{\end{equation}}
\newcommand{\bes}{\begin{equation}\begin{split}}

\def \d {{\rm d}}

\definecolor{darkred}{rgb}{0.9,0,0}

\begin{document}
\vspace{-5.0cm}
\begin{flushright}
OUTP-20-11P,
TTP20-038, P3H-20-065
\end{flushright}

\vspace{2.0cm}

\begin{center}
{\Large \bf
On the non-cancellation of infrared singularities in collisions of massive quarks }\\
\end{center}

\vspace{0.5cm}

\begin{center}
Fabrizio Caola$^{1,2}$,
Kirill Melnikov$^{3}$, Davide Napoletano$^{4}$, Lorenzo Tancredi$^{1}$\\
\vspace{.3cm}
{\it
{}$^1$Rudolf Peierls Centre for Theoretical Physics, Clarendon Laboratory, Oxford, UK\\
{}$^2$Wadham College, University of Oxford, Oxford OX1 3PN, UK\\
{}$^3$Institute for Theoretical Particle Physics, KIT, Karlsruhe, Germany\\
{}$^4$Universit\`a Milano-Bicocca \& INFN, Piazza della Scienza 3, Milano 20126, Italia
}

\vspace{1.3cm}

{\bf \large Abstract}
\end{center}
We discuss  the infrared structure of processes with  massive quarks in the initial state.
 It is well known that,  starting from next-to-next-to-leading order
 in perturbative QCD, such  processes exhibit a violation  of the Bloch-Nordsieck
 theorem, in that the
 sum of real and virtual contributions to partonic cross sections
 contains uncanceled infrared
singularities. The main purpose of this paper is to present a simple physical argument
that elucidates  the origin of these singularities and simplifies   the derivation of
 infrared-singular  contributions to heavy-quark initiated cross sections.



\pagenumbering{arabic}

\allowdisplaybreaks

\section{Introduction and general considerations}
The infrared structure of perturbative gauge theories is a fascinating
topic  which  received significant  attention since the early days of
QCD~\cite{Yennie:1961ad,KLN}.  In the 70`s, the observation  of factorization of
soft and collinear divergences in deep-inelastic
scattering~\cite{Georgi-Politzer,Gross-Wilzceck} paved the way for a
new understanding of the perturbative structure of gauge theories,
leading  to the promotion of the naive parton model~\cite{feynman}
to a well-defined approximation rooted in a  fully consistent quantum theory
of strong interactions.

Generalisation of  these results to the more complicated case of
hadron-hadron collisions~\cite{keith&friend,DGLAP,Libby-Sterman,CSS}  resulted in a
better understanding on the universal pattern of factorization and
cancellation of long-distance effects in perturbative QCD
calculations.  This  understanding was   eventually distilled
into ``theorems'' ~\cite{collins,foundation}
that state that (potential) logarithmic sensitivity to long-distance effects is absent in  sufficiently inclusive
observables in hard scattering processes. This remarkable fact is the foundation of modern collider phenomenology
as it allows us to to provide {\it first-principles}  improvements
of the theoretical description of hadron  collisions by refining  predictions for  {\it partonic} scattering
cross sections in QCD perturbation theory.

Given the prominence of these theorems in modern collider physics, it
is useful to inquire  about their limitations. Such a question, albeit  being
interesting in its own right,  may also have
practical consequences  for the precision physics program at current and future
colliders by e.g. informing  us about   ultimate limits in precision
that improvements in  perturbative computations {\it alone}
can  possibly provide.

Indeed, while the aforementioned theorems are very solid in the case of
lepton-lepton or lepton-hadron collisions, the situation is more
delicate in case of  hadron-hadron collisions, see e.g.~\cite{collins,
foundation}.
In fact, it was
argued  that, at sufficiently high  orders in perturbation theory,
combining  real and virtual corrections within the framework of collinear
factorization may be insufficient to get rid of the infrared sensitivity, even for inclusive
observables~\cite{seymour,andrzej,catani}.

For processes involving massless
partons in the initial state, our current understanding of the
soft-collinear structure of QCD implies that these issues  can only
appear at third or higher orders in QCD perturbation theory.\footnote{At third
order, they are only relevant for processes involving a non-trivial
color structure~\cite{catani,lance,claude}.}
However, the situation is  very different if one considers  massive quarks
in the initial state. In this case, it was pointed out long ago
that starting from  second order in QCD perturbation theory
the sum of real and virtual corrections is not free of
infrared singularities. As a
consequence, ``standard'' perturbative calculations in this case become insufficient beyond
next-to-leading order, even for the simplest partonic processes~\cite{doriamassive}.

This problem received a lot of attention in the
past~\cite{doriacataniold} and several formal ways of dealing with it
have been proposed~\cite{mutaetal}.  The goal of this paper is to
present a derivation of the divergent contribution to the cross
section of a process with two heavy quarks in the initial state that,
in our opinion, is remarkably simple and physically transparent.

Our argument is inspired by
recent work on  infrared subtraction schemes for higher order
calculations \cite{czakonfks,nested} 
and, in a nutshell, consists in connecting  infrared singular contributions of a process
where infrared finiteness is guaranteed to infrared singular contributions of a process initiated by the collision
of two massive quarks.
In what follows, we  focus on the Drell-Yan process where a virtual photon is produced in the collision
of a quark and an anti-quark. The simplicity
of this process allows us to present our argument with a minimal amount of technical overhead.

The remainder of this paper is organised as follows.
In Sec.~\ref{sec:nlo}
we show by an explicit computation that there are no uncanceled infrared singularities at
next-to-leading order (NLO) QCD for the Drell-Yan process with massive initial-state quarks and
comment on the generalization of this result to arbitrary processes. We also argue that the absence of
infrared singularities in the production process $q \bar q \to V +X$ at NLO QCD
can be naturally understood if the absence  of infrared singularities in the decay process $ V \to q \bar q + X$
is taken for granted.   In Sec.~\ref{sec:nnlo},
we generalize this  argument to the next-to-next-to-leading order (NNLO) case and show that at NNLO there is
only one potential source of  non-canceling soft singularities. In
Sec.~\ref{sec:one-loop}, we explicitly compute the infrared  singular  contribution to the Drell-Yan cross section
and comment on the result. We
conclude in Sec.~\ref{sec:conclusions}. The  analytic
continuation of the one-loop integrals required for our analysis
is discussed  in the  appendix.

\section{Drell-Yan process with initial-state massive quarks}

We begin with the discussion of the infrared structure of the process
\be
q(p_1)+\bar q(p_2) \to V(p_V) + X,
\label{eq1}
\ee
where $q,\bar q$ are massive  quarks with $p_1^2 = p_2^2 = m_q^2$
and $V$ is a virtual photon\footnote{
  Our argument applies verbatim for  any (massive) color-singlet
  final state $V$.} with $p_V^2=m_V^2$. Since there are no massless partons in the initial state of this process,
   no collinear renormalization of parton distribution functions is required.
   The perturbative expansion of the  partonic cross section for this process
   reads
   \be
\d\sigma = \d\sigma_{\rm LO} + \d\sigma_{\rm NLO} + \d\sigma_{\rm NNLO}+
\mathcal O(\alpha_s^3).
\ee

\subsection{Next-to-leading order}
\label{sec:nlo}
We start by considering next-to-leading order (NLO) QCD contributions  to the cross section of the process
in Eq.~\eqref{eq1}.  We write them as
\be
\d\sigma_{\rm NLO} = \d\sigma_{\rm V} + \d\sigma_{\rm R}.
\label{eq:nlo}
\ee
The first term on the r.h.s. of Eq.~\eqref{eq:nlo} represents UV-renormalized contributions of one-loop
virtual
corrections. It reads~\cite{Catani:2000ef}
\be
\d\sigma_{\rm V} = \frac{\alpha_s(\mu)}{2\pi}
\left\{-\frac{2 C_F}{\ep}
  \left[
  \frac{1}{2v}\ln\lp\frac{1-v}{1+v}\rp+1
  \right]
  \right\}\d\sigma_{\rm LO} + \d\sigma_{\rm V,fin},
  \label{eq:sigmaV}
\ee
where  $\ep = (4-d)/2$ and $d$ is the dimensionality of space-time.
Also,
$C_F = 4/3$ is the Casimir invariant of the $SU(3)$ gauge group of QCD,
$
v = \sqrt{1-m^4/(p_1\cdot p_2)^2} 
$
and $\d\sigma_{\rm V,fin}$ is finite in the $\ep \to 0$ limit.
The $1/\ep$ pole in Eq.~\eqref{eq:sigmaV} is of infrared origin; it is well known
that it is  canceled by  a similar divergence  in the real emission
contribution  $\d\sigma_{\rm R}$.

To illustrate  this,  consider the real emission process\footnote{
  We only consider the corrections to the $q\bar q$ channel, since the $qg$
  channel is infrared finite.}
\be
q(p_1)+\bar q(p_2)\to V(p_V) + g(p_g),
\label{eq:rproc}
\ee
and write
\be
\d\sigma_{\rm R} = \frac{1}{4J} \int [{\rm d} p_V] [{\rm d} p_g]
\overline\sum
| \mathcal M_0 (p_1,p_2; p_V,p_g)|^2
(2\pi)^d \delta_{d}(p_1+p_2-p_V-p_g),
\label{eq:nlor}
\ee
where $J = p_1 \cdot p_2 \; v $  is the flux factor,  $[{\rm d} p_{V,g}] = \d^{d-1}p_{V,g} /( (2\pi)^{d-1} 2 E_{V,g})$
are the phase-space elements of the virtual photon
and the gluon, respectively,
$\overline \Sigma$ indicates the
sum (average) over final-state (initial-state)  colors and polarizations, and
$\mathcal M_0$ is the tree-level
scattering amplitude for  the process Eq.~\eqref{eq:rproc}.
When the emitted gluon becomes soft, $E_g \to 0$, the matrix
element $|\mathcal M_0|^2$ scales as $E_g^{-2}$, and  Eq.~\eqref{eq:nlor} develops
a logarithmic singularity. To expose it, we work in the partonic center-of-mass frame, separate the
integration over the gluon energy and  write
\be
 \d\sigma_{\rm R} =\frac{1}{4J}
\int  \frac{ {\rm d} E_g}{E_g^{1+2\ep} } \frac{\d\Omega^{(d-1)}_{g}}{2 (2\pi)^{d-1}} \;
F^{(d)}_{g}(p_1,p_2,p_V; p_g),
\label{eq2.8}
\ee
where
\be
F^{(d)}_{g}(p_1,p_2,p_V; p_g ) =  \frac{1}{4J}\; [{\rm d} p_V]
E_g^2\;
\overline\sum
| \mathcal M_0 (p_1,p_2; p_V,p_g)|^2 \;
(2\pi)^d \delta_{d}(p_1+p_2-p_V-p_g).
\ee

To extract infrared divergences from Eq.~\eqref{eq2.8}, we write
\be
 \d\sigma_{\rm R} =
\int \limits_{0}^{\rm E_{\rm max}} \frac{ {\rm d} E_g}{E_g^{1+2\ep} } \frac{\d\Omega^{(3)}_{g}}{16\pi^3} \;
\lim_{E_g \to 0} \left [ F^{(4)}_{g}(p_1,p_2,p_V; p_g) \right ]  + \d\sigma_{\rm R}^{\rm fin},
\ee
where the second contribution is finite and the first one is divergent.  We rewrite it as
\be
\int \limits_{0}^{\rm E_{\rm max}} \frac{ {\rm d} E_g}{E_g^{1+2\ep} } \frac{\d\Omega^{(3)}_{g}}{16\pi^3} \;
\lim_{E_g \to 0} \left [ F^{(4)}_{g}(p_1,p_2,p_V; p_g)  \right ] = {\rm d} \sigma_R^{\rm div} + \cdots,
\label{eq2.11}
\ee
where
\be
{\rm d} \sigma_{\rm R}^{\rm div}=
-\frac{1}{2\ep} \int \frac{\d\Omega^{(3)}_{g}}{16\pi^3} \lim \limits_{ E_g \to 0}^{}
\left [ F^{(4)}_{g}(p_1,p_2,p_V; p_g) \right ],
  \label{eq:almost}
\ee
and the ellipses in Eq.~\eqref{eq2.11} stand for finite terms.

To proceed further, we recall that in the soft limit scattering amplitudes
obey the well-known factorization formula
\be
\mathcal M_0(p_1,p_2;p_V,p_{g^a})\approx g_s^2 \varepsilon^\mu
J_\mu^{a,(0)}(p_1,p_2;p_g) \mathcal M_0(p_1,p_2;p_V),
\label{eq:fact_tree}
\ee
where $\varepsilon^\mu$ is the gluon polarization vector and  $a$ is its color
index.  The tree-level soft current reads
\be
J_\mu^{a,(0)}(p_1,p_2;p_g) = \sum_{i=1}^{2} T_i^a \frac{p_{i,\mu}}{p_i\cdot p_g},
\label{eq:softtree}
\ee
where  $T_i^a$ is the color charge of  particle
$i$. In our case, $T^a_1 = t^a_{21}$ and $T^a_2 = -t^a_{12}$, where $t^a_{ij}$
is the  matrix element of an $SU(3)$ algebra generator in the fundamental representation.\footnote{For
  more details on the color notation, see e.g.~\cite{cataniseymour}.}
This immediately allows us to rewrite Eq.~\eqref{eq:almost} as
\be
\d\sigma_{\rm R}^{\rm div} = {\rm Eik}_{0}(p_1,p_2)\times \d\sigma_{\rm LO},
\ee
where
\be
{\rm Eik}_{0}(p_1,p_2) =
-\frac{\alpha_s(\mu)}{2\pi}
\frac{C_F}{\ep} \int \frac{\d\Omega_{3,g}}{4\pi} E_g^2\left[
  \frac{2(p_1\cdot p_2)}
       {(p_1\cdot p_g)(p_2\cdot p_g)}-
       \frac{m_q^2}{(p_1\cdot p_g)^2}-
       \frac{m_q^2}{(p_2\cdot p_g)^2}
       \right].
\label{eq:eiknlo}
\ee
We parametrise  momenta in Born kinematics as
$p_{1,2} = m_V/2\;(1,0,0,\pm\beta)$, with
$\beta = \sqrt{1-4m_q^2/m_V^2}$ and
$p_g = E_g(1,\sin\theta,0,\cos\theta)$. A straightforward integration over the gluon emission angle
leads to
\be
\d\sigma_{\rm R}^{\rm div} = \frac{\alpha_s(\mu)}{2\pi}
\times
\frac{2C_F}{\ep} \left[
  \frac{1+\beta^2}{2\beta}\ln\lp\frac{1-\beta}{1+\beta}\rp+1
  \right]\d\sigma_{\rm LO}.
\label{eq:sigmaRfinal}
\ee
The cancellation of soft singularities in the NLO cross section can be observed upon combining
$\d\sigma_{\rm V}$ from Eq.~\eqref{eq:sigmaV} and $\d\sigma_{\rm R}^{\rm div}$ from Eq.~\eqref{eq:sigmaRfinal}
and using  the relation between  $v$ and $\beta$,  $v =
2\beta/(1+\beta^2)$, which implies
\be
\frac{1+\beta^2}{2\beta}\ln\lp\frac{1-\beta}{1+\beta}\rp =
\frac{1}{2 v} \ln\lp\frac{1-v}{1+v}\rp.
\ee

We also  note that the cancellation of infrared divergences occurs  in a much broader context than what
we discuss here for the Drell-Yan process.  Indeed,  by considering a generalization of
Eq.~\eqref{eq:sigmaV} to $2 \to n$ processes as described in  Ref.~\cite{Catani:2000ef},
and adapting Eq.~\eqref{eq:softtree} to this case, it is straightforward to prove the
cancellation of infrared divergences for arbitrary processes with massive quarks in the initial state.

We will now re-analyse the NLO case from a perspective that will be helpful for deriving the infrared divergent contribution
to the NNLO cross section.  To this end, instead
of considering  the production process $q \bar q \to V+X$, we  start with its decay
counterpart  $V(p_V) \to q(p_1) + \bar q(p_2) + X$.
We use the optical theorem and obtain  the total decay rate of the above process  from the imaginary
part of the  time-ordered
correlator of two vector currents. Since such
correlator cannot have infrared divergences, we conclude that the
decay rate is  free of infrared singularities as well.
Writing the  decay rate as  the sum of virtual and real-emission  contributions, we conclude that
$\d\sigma_{\rm V}^{\rm decay} + \d\sigma_{\rm R}^{\rm decay}$  is infrared finite.

We now want to relate $\d\sigma_{\rm V}^{\rm decay}$,
$\d\sigma_{\rm R}^{\rm decay}$ to their counterparts in  the production case
Eqs~(\ref{eq:sigmaV}, \ref{eq:nlor}). For virtual corrections, this relation
is obvious. Indeed, one-loop corrections to the $\gamma^* \to q \bar q $   vertex are
described by a  single form factor $F_V$ that {\it only} depends on the invariant mass of the
virtual photon $m_V^2$.\footnote{The dependence of the form factor on quark masses is not relevant for this discussion.}
Hence, this form factor
is identical for the production ($q \bar q \to \gamma^*$) and decay ($\gamma^* \to q \bar q$)
processes.  We conclude  that
the infrared structure of the decay rate $\d\sigma_{\rm V}^{\rm decay}$ and the production cross section
$\d\sigma_{\rm V}$, is the same. Therefore
\be
\begin{split}
 \d\sigma_{\rm V}^{\rm decay} = F_{V}(m_V^2,\ep)\d\sigma_{\rm LO}^{\rm decay} + \cdots
 ~{\rm and}~~~\d\sigma_{\rm V} = F_{V}(m_V^2,\ep)\d\sigma_{\rm LO} + \cdots,
 \label{eq2.18}
\end{split}
\ee
where the ellipses stand for finite contributions.

To make use of the  finite nature of NLO corrections to the  decay as an explanation of  why NLO corrections
to the production are  finite,  we need to understand how the real
emission contribution to the decay rate  changes when  we move  heavy quarks  into  the initial state
and the vector boson into  the final state which is required for calculating the production cross section.
Since we are only interested in the divergent
contribution to the cross section, we require  this crossing in the soft limit.
We note that the tree-level
soft current Eq.~\eqref{eq:softtree} is homogeneous in the hard momenta
$p_{1,2}$ (and it does not depend on the momenta of the color singlet), so
it does not change under the replacement  $p_i\to -p_i$. Moreover,
the phase space of the Born  process decouples from the eikonal factor and the gluon
phase space in the soft limit. It follows that
\be
\begin{split}
  \d\sigma^{\rm decay}_{\rm R} =
          {\rm Eik}_{0}(p_1,p_2)\times \d\sigma^{\rm decay}_{\rm LO}
          + \cdots
          ~~~{\rm and}~~~
          \d\sigma_{\rm R} =
                  {\rm Eik}_{0}(p_1,p_2)\times \d\sigma_{\rm LO} + \cdots,
\end{split}
\label{eq2.19}
\ee
where ellipses stand for finite contributions and
the function  ${\rm Eik}_0$ is defined in Eq.~\eqref{eq:eiknlo}.
Since $\d\sigma^{\rm decay}_{\rm V} + \d\sigma^{\rm decay}_{\rm R}$ is free of infrared divergences,
it  follows from Eqs~(\ref{eq2.18}, \ref{eq2.19}) that
\be
F_V(m_V^2,\ep) + {\rm Eik}_{0}(p_1,p_2)
\ee
is infrared finite.  Without any additional computation, this ensures that the ${\cal O}(\alpha_s)$ contributions
to the
cross section of $q \bar q \to V+X$ with massive initial state quarks are finite as well.
In the next section we generalize this analysis to next-to-next-to-leading order.

\subsection{Next-to-next-to-leading order contributions to the production cross section}
\label{sec:nnlo}

Consider the NNLO QCD contributions to the cross section of the production process $q \bar q \to V + X$.
In full analogy to the NLO case discussed in the previous section, we split $\d\sigma_{\rm NNLO}$ into
double-virtual, double-real and real-virtual contributions
\be
\d\sigma_{\rm NNLO} = \d\sigma_{\rm VV} + \d\sigma_{\rm RR} + \d\sigma_{\rm RV}.
\ee
In this equation, the double-virtual term $\d\sigma_{\rm VV}$ is proportional
to the two-loop form
factor for the $q\bar q\to V$ process.
The double-real term $\d\sigma_{\rm RR}$ is proportional to the
tree-level matrix element for the process
\be
q(p_1) + \bar q(p_2) \to V(p_V) + f_i(p_i) + f_j(p_j),
\ee
where $(f_j,f_j)\in \{(g,g),(q_i,\bar q_j)\}$ and $q_i$ is a generic
(massive or massless) quark. Finally, the real-virtual contribution
$\d\sigma_{\rm RV}$ is
proportional to the one-loop matrix element for the process
\be
q(p_1) + \bar q(p_2) \to V(p_V) + g(p_g).
\ee

In principle, one can study the infrared structure of the various contributions at this perturbative
order by  extending the NLO analysis presented at the beginning of the previous section to one order higher.
However, it is much easier and more transparent
to re-use the connection between the production and decay processes  as was done at the end
of the previous section.  For this reason,  we consider the NNLO QCD  contributions to the decay process
$ V \to q \bar q + X$, which is finite, 
and write
\be
\d\sigma_{\rm NNLO}^{\rm decay} =
\d\sigma_{\rm VV}^{\rm decay} +
\d\sigma_{\rm RR}^{\rm decay} +
\d\sigma_{\rm RV}^{\rm decay}.
\ee
We then  compare each contribution to its counterpart in  the production case.
The results of this comparison can be summarized as follows.
\begin{itemize}
\item All infrared singularities of the double-virtual contributions  come
  from the one- and two-loop $Vq\bar q$ form factors. Since the form factor is
  the same for the $V\to q+\bar q$ and $q+\bar q\to V$ processes, the infrared
  structure of $\d\sigma_{\rm VV}^{\rm decay}$ and $\d\sigma_{\rm VV}$ is
  identical.
\item In the double-real contribution, infrared singularities appear when either one or two
	final state gluons become soft, or when a massless final state quark pair becomes soft.
	The case of one-gluon emission is described by the tree-level current Eq.~\eqref{eq:softtree}.
	The emission of two soft partons is described  by a double-soft current~\cite{Catani:1999ss} that
         is homogeneous in the
  momenta of the external hard partons.  Similar to the NLO case described above,
	this implies that the infrared structure of
  $\d\sigma_{\rm RR}^{\rm decay}$ and $\d\sigma_{\rm RR}$ is identical.
 \item The real-virtual contribution contains both explicit $1/\ep$ infrared poles in the
$q\bar q\to V + g$ one-loop amplitude and implicit singularities that only appear
after integrating over the soft region of the  gluon phase space. As long as the gluon
is hard, this integration does not introduce any divergence and
only explicit singularities are relevant. These singularities  cancel
against single soft-gluon emission in the double-real contribution
along the lines of the  NLO case described in the previous section.
As we explained there,
this cancellation occurs for both   the production and the decay processes.

The only contribution
that we still need to discuss is a one-loop correction to the emission of a  \emph{soft} gluon.
In this case, we cannot  invoke the crossing argument to conclude that
the production and decay processes share the same infrared structure because the analytic structure of
loop amplitudes is non-trivial and  care is needed to relate the production and decay cases.
\end{itemize}

Hence, we conclude that the infrared structure of the production and decay processes
is identical, \emph{except for  possible contributions that originate  from crossing the $V\to q+\bar q + g$
one-loop amplitude into the $q+\bar q \to V + g$ one, in the kinematic configuration where $g$ is soft.}
\emph{ Since the total rate for $V\to q+\bar q+X$ is finite, this implies that the only potential non-canceling infrared
singularities in $q+\bar q\to V+X$ at NNLO must  be related to this crossing. } Below  we
show that  the analytic continuation from decay to production kinematics is indeed non-trivial, and that
it leads to an   uncanceled $1/\ep$ infrared singularity in the production cross section.\footnote{
We note that similar arguments suggest that  other partonic channels, i.e.  $qg$ and $gg$, are infrared-finite.}

\section{The one loop soft current and its crossing}
\label{sec:one-loop}
In this section, we study one-loop corrections to soft gluon emission. More precisely, following the discussion
in the previous section,  we investigate whether {\it additional}  infrared divergences
can be generated by crossing the one-loop decay amplitude $\mathcal M_1(p_V;p_1,p_2,p_g)$ into
the amplitude $\mathcal M_1(p_1,p_2;p_V,p_g)$ that describes the production process.

Similar to the tree-level case Eq.~\eqref{eq:fact_tree},
the one-loop amplitude $\mathcal M_1$ also factorizes in the soft
limit\footnote{In this equation, $g_s$ is the bare strong coupling. Since we are interested in
infrared effects, we do not discuss renormalization.}
\be
\begin{split}
\mathcal M_1 (p_V;p_1,p_2,p_g) \approx g_s^2 \varepsilon^\mu
\bigg[ J_\mu^{a,(0)}(p_1,p_2;p_g) \mathcal M_1 (p_V;p_1,p_2)
\\+
g_s^2 J_\mu^{a,(1)}(p_1,p_2;p_g) \mathcal M_0 (p_V;p_1,p_2)\bigg].
\end{split}
\label{eq:m1soft}
\ee
We stress that $\mathcal M_1$ in the above equation is the scattering amplitude of the decay process and we
intend to get the production amplitude by crossing.

The tree-level current $J_\mu^{a,(0)}$ is given in Eq.~\eqref{eq:softtree};  as discussed in
Sections~\ref{sec:nlo}, \ref{sec:nnlo} it leads to the same infrared divergences in  the production
and decay cases.  Hence, we only need to focus on the second term on the right hand side of
Eq.~\eqref{eq:m1soft} that describes the one-loop correction to the soft current.

\begin{figure}
\centering
\includegraphics[width=0.95\textwidth]{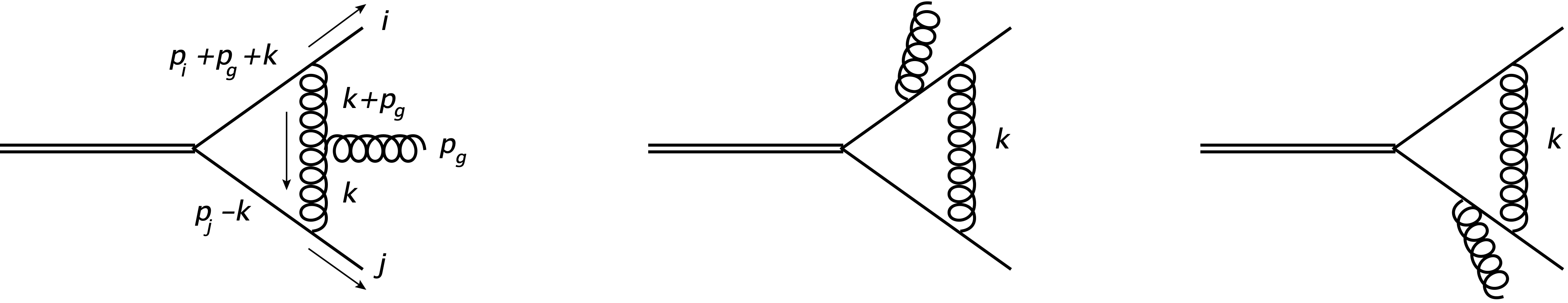}
\caption{Diagrams contributing to the one-loop soft current.
$i$ and $j$ are hard eikonal
lines, $p_g$ and $k$ are soft, see text for details. }
\label{fig:soft}
\end{figure}

To compute the one-loop soft current $J_\mu^{a,(1)}$, one needs to consider the non-abelian part of the
diagrams shown in Fig.~\ref{fig:soft}, in the limit where both virtual and real gluons
are soft~\cite{Catani:2000pi}. The result reads
\be
\begin{split}
J^{a,(1),\mu}(p_1,p_2;p_g) &=
i f_{abc}\sum_{\substack{i,j=1\\i \neq j}}^2 T_i^b T_j^c
\lp\frac{p_i^\mu}{p_i\cdot p_g}-\frac{p_j^\mu}{p_j\cdot p_g}\rp
g_{ij}^{(1)}(\ep,p_g;p_i,p_j)\\
&=
g_{12}^{(1)}(\ep,p_g;p_1,p_2) \;C_A\; J^{a,(0),\mu}(p_1,p_2;p_g).
\end{split}
\label{eq:softloop}
\ee
where $f_{abc}$ are the $SU(3)$ structure constants and $g_{ij}^{(1)}$ is a function that will
be specified later.  We stress that $J^{a,(1)}_\mu$ is purely
non-abelian. This feature is expected because
in an abelian theory the tree-level soft current does not receive corrections.
Since, as we argued at the beginning of this section,
Eq.~\eqref{eq:softloop} provides the  {\it only} source of non-canceling soft singularities for the process
$q\bar q \to V+X$ with massive initial particles, we
recover  the classic result that
\emph{in the abelian (e.g. QED) case the NNLO cross-section for the collision of
  two massive partons is infrared-finite.}

We continue with
the non-abelian case. Following the argument of Sec.~\ref{sec:nnlo}, we investigate whether Eq.~\eqref{eq:softloop} leads to the
same infrared structure for the decay and production processes.
Since $J^{a,(0),\mu}$ is invariant under  $p_{1,2}\to -p_{1,2}$, any potential difference must come
from the crossing of $g_{12}^{(1)}$. It is easy to see that,  at NNLO,
only the real part of $g_{12}^{(1)}$ contributes to the cross section; for this reason we investigate
the behavior of $\Re\big[g_{12}^{(1)}\big]$ under $p_{1,2}\to-p_{1,2}$ transformation.

It is instructive to consider first the case of massless  quarks. For $m_q=0$,
the function $g_{12}^{(1)}$ reads~\cite{Catani:2000pi}
\be
g_{12}^{(1)}(\ep,p_g;p_1,p_2) =
-\frac{1}{16\pi^2}\frac{1}{\ep^2}
\frac{\Gamma^3(1-\ep)\Gamma^2(1+\ep)}{\Gamma(1-2\ep)}
\left[\frac{(-s_{12}-i\delta)}{(-s_{1g}-i\delta)(-s_{2g}-i\delta)}\right]^{\ep},
\label{eq3.3}
\ee
with  $s_{ij} = 2 p_i\cdot p_j$. This implies
\be
\Re\left[
g_{12}^{(1)}(\ep,p_g;-p_1,-p_2)\right] = \Re\left[g_{12}^{(1)}(\ep,p_g;p_1,p_2)\right].
\ee
The argument of Sec.~\ref{sec:nnlo} then allows us to reproduce the standard result that for massless
quarks the cross section for the process $q+\bar q\to V$ is free from soft singularities at NNLO.\footnote{To
  remove initial-state collinear singularities, one still needs  to redefine  parton distribution functions in the case
of massless particles collisions.}

We  continue with the case $m_q\ne 0$.  In this case, we follow
Ref.~\cite{czakon} and write $g_{12}^{(1)}$ as
\be
g_{12}^{(1)}(\ep,p_g;p_1,p_2) = \sum_{i=1}^{3}f_i(p_g;p_1,p_2) M_i(\ep,p_g;p_1,p_2),
\label{eq:gMdef}
\ee
where $M_i$ are defined as
\begin{align}
&M_1(\ep,p_g;p_1,p_2) = \int \frac{d^d k}{(2\pi)^d} \frac{1}{[k^2+i\delta][(k+p_g)^2+i\delta][-2p_2\cdot k + i \delta]},
\notag
\\
&M_2(\ep,p_g;p_1,p_2) = \int \frac{d^d k}{(2\pi)^d} \frac{1}{[k^2+i\delta][2p_1\cdot k + 2 p_1\cdot p_g + i \delta]
[-2p_2\cdot k + i \delta]},
\label{eq:Mdef}
\\
&M_3(\ep,p_g;p_1,p_2) = \int \frac{d^d k}{(2\pi)^d} \frac{1}{[k^2+i\delta][(k+p_g)^2+i\delta][2p_1\cdot k + 2 p_2\cdot p_g + i \delta]
[-2p_2\cdot k + i \delta]},
\notag
\end{align}
and $f_i$ are rational functions of $p_i\cdot p_j,p_i\cdot p_g$.
Since $g_{12}^{(1)}$ has to be computed using
eikonal vertices~\cite{Catani:2000pi}, it follows that
\be
f_1(p_g;-p_1,-p_2) = -f_1(p_g;p_1,p_2),~~~~~
f_{2,3}(p_g;-p_1,-p_2) =  f_{2,3}(p_g;p_1,p_2).
\label{eq:fdef}
\ee
The explicit form of $f_i$ can be found in Ref.~\cite{czakon}, but it is not needed for our argument. 

Using Eqs~(\ref{eq:gMdef}, \ref{eq:Mdef}, \ref{eq:fdef}) one can show by analytic continuation of the $M_j$ integrals 
 that
the function $g_{12}^{(1)}$ changes in the following way
\be
g_{12}^{(1)}(\ep,p_g;-p_1,-p_2) = e^{-2i\ep\pi} g_{12}^{(1)}(\ep,p_g;p_1,p_2).
\label{eq:cross}
\ee
This is worked out explicitly in Appendix~\ref{sec:app}. 
To proceed further,  we  write the (decay) function $g_{12}^{(1)}$  as
\be
g_{12}^{(1)}(\ep,p_g;p_1,p_2) = \frac{\alpha_s}{2\pi}
E_g^{-2\ep}
\sum_{k=-2}^{\infty} \left[
\mathfrak r_{k} + i\cdot \mathfrak i_{k}
\right]\ep^k,
\label{eq:param}
\ee
with $\mathfrak r$ and $\mathfrak i$ real and $\mathfrak i_{-2} = 0$, see Appendix~\ref{sec:app}.
Using Eqs ~(\ref{eq:cross}, \ref{eq:param}) we can then write the difference between the real
 parts of the functions $g_{12}$ required to describe the production and the decay processes
as
\be
\begin{split}
&\Re\bigg[g_{12}^{(1)} (\ep,p_g;-p_1,-p_2)\bigg] -
\Re\bigg[g_{12}^{(1)} (\ep,p_g;p_1,p_2)\bigg]
\\
& =
\frac{\alpha_s}{2\pi}\left|\frac{s_{12}}{s_{1g}s_{2g}}\right|^\ep\big[
-2\pi^2 \cdot \mathfrak r_{-2} +
2\pi\cdot \mathfrak i_{-1} + \mathcal O(\ep)\big].
\end{split}
\label{eq:eqcont}
\ee
Since the real part of  $g_{12}^{(1)}$ at order ${\cal O}(\ep^0)$ contributes to divergences of the cross section or decay rate
at order $1/\ep$,
the argument presented in Sec.~\ref{sec:nnlo}  implies that \emph{the second line
of Eq.~\eqref{eq:eqcont} gives rise to a non-canceling
infrared divergence in the NNLO cross section for the $q+\bar q\to V$ process} with massive  quarks in the initial
state.

This non-canceled singularity is controlled by the coefficients $\mathfrak r_{-2}$ and
$\mathfrak i_{-1}$. They can be immediately obtained by matching Eq.~\eqref{eq:m1soft} to   the universal
expression for the infrared poles of one-loop amplitudes~\cite{Catani:2000ef}. We obtain
\be
\mathfrak r_{-2} = -\frac{1}{2},~~~~\mathfrak i_{-1}=\pi\left(\frac{1}{2v}-1\right),
\ee
with $v$ defined immediately after Eq.~\eqref{eq:sigmaV}.
 We work in the center of mass frame of the two quarks and
rewrite Eq.~\eqref{eq:eqcont} as
\be
\begin{split}
&\Re\bigg[g_{12}^{(1)} (\ep,p_g;-p_1,-p_2)\bigg] =
\Re\bigg[g_{12}^{(1)} (\ep,p_g;p_1,p_2)\bigg]
+\frac{\alpha_s}{2\pi} E_g^{-2\ep}
\left[
\lp\frac{1-v}{v}\rp\pi^2+\mathcal O(\ep)\right].
\end{split}
\label{eq:mismatch}
\ee
To find the contribution of the last term in Eq.~\eqref{eq:mismatch} to the cross section, we note that
the soft current at one loop  is proportional to the tree-level one, cf.  Eq.~\eqref{eq:softloop}.  As a consequence,
we can read off the required result directly from  Eq.~\eqref{eq:sigmaRfinal} that describes the NLO calculation.\footnote{Note that
the additional $E_g^{-2\ep}$ factor in Eq.~\eqref{eq:mismatch} would give rise to an extra factor 1/2
compared to the NLO case. This is compensated however by the factor of 2 in $2\Re[\mathcal M_0 \mathcal M_1^{*}]$.}
Therefore, we write the real-virtual contribution to the decay process as
\be
\d\sigma_{\rm RV}^{\rm decay} = {\rm Eik}_1(p_1,p_2) \times \d\sigma_{\rm LO}^{\rm decay} + \cdots,
\label{eq:rvfin1}
\ee
where the ellipses stand for finite contributions. The real-virtual contribution to the production
process is given by
\be
\d\sigma_{\rm RV} = 
{\rm Eik}_1(-p_1,-p_2) \times \d\sigma_{\rm LO} =
{\rm Eik}_1(p_1,p_2) \times \d\sigma_{\rm LO} +
 \Delta [ \d\sigma_{\rm RV}^{\rm div}] + \cdots .
\label{eq:rvfin2}
\ee
The second term in the r.h.s. of Eq.~\eqref{eq:rvfin2} is the {\it additional} divergent contribution to the production
cross section caused by a non-trivial  analytic continuation of soft loop integrals upon crossing.  It reads
\be
\Delta [ \d\sigma_{\rm RV}^{\rm div} ] = \left[\frac{\alpha_s(\mu)}{2\pi}\right]^2
\frac{2 C_A C_F\;\pi^2}{\ep} \left[\frac{1}{2v}\ln\lp\frac{1-v}{1+v}\rp +1\right]
\lp\frac{1-v}{v}\rp\d\sigma_{\rm LO}.
\label{eq:IR}
\ee
Thanks to  the argument
presented in Sec.~\ref{sec:nnlo}, we conclude that the cross section for $q\bar q\to V+X$ with massive
quarks in the initial state  contains non-canceling infrared divergence given by $\Delta [ \d\sigma_{\rm RV}^{\rm div} ]$.
Therefore,
\be
\begin{split}
\d\sigma_{\rm NNLO} & = \Delta [ \d\sigma_{\rm RV}^{\rm div} ] + \cdots =
\\\
& \left[\frac{\alpha_s(\mu)}{2\pi}\right]^2
\frac{2 C_A C_F\;\pi^2}{\ep} \left[\frac{1}{2v}\ln\lp\frac{1-v}{1+v}\rp +1\right]
\lp\frac{1-v}{v}\rp\d\sigma_{\rm LO}+ \cdots,
\label{eq3:16}
\end{split}
\ee
where the ellipses  stand for finite contributions to the NNLO cross section.  Eq.~(\ref{eq3:16})  describes
the violation
of Bloch-Nordsieck cancellations~\cite{BN}  in the case when two massive quarks  collide. It coincides
with the expression derived in Refs.~\cite{doriamassive,doriacataniold,mutaetal}.

We now comment on the result Eq.~\eqref{eq3:16}. First, we note that in the massless case $v\to 1$ and
  the  divergence disappears.   A simple generalization of this result
  to the collision of two quarks with unequal masses  shows that Eq.~\eqref{eq3:16} remains valid  provided that
  $v = \sqrt{1 - m_1^2 m_2^2/(p_1 p_2)^2}$.  It follows that the divergence in Eq.~(\ref{eq3:16})
  disappears if  only one quark in the initial state is massive.

Moreover, Eq.~\eqref{eq:IR}  implies that the non-canceling infrared divergences  in cross sections
with massive quarks in the initial state are power-suppressed
\be
\Delta [ \d\sigma_{\rm RV}^{\rm div} ]  \sim  \mathcal O\lp \frac{m_q^4}{m_V^4}  \rp {\rm d} \sigma_{\rm LO}.
\label{eq3.14}
\ee
This behavior is compatible with classic arguments about factorization, see e.g. Ref.~\cite{ESW} for a review.
In fact, a small mass of the quark in the initial state probes the sensitivity of the partonic cross section
to long-distance physics.  The result Eq.~\eqref{eq3.14} then informs us that at the level of {\it logarithmic}
sensitivity to long-distance effects, the partonic cross section is certainly infrared finite.
The non-cancellation of infrared divergences at the level of ${\it power~corrections}$,
as indicated in Eq.~\eqref{eq3.14}, simply shows that an understanding
of factorization for  higher-twist or, in general, power corrections is required to make
calculations with massive partons self-consistent.

\section{Conclusion}
\label{sec:conclusions}
It is well known \cite{doriamassive,doriacataniold,mutaetal} 
that  partonic cross sections computed with massive quarks in the initial state are not
infrared finite starting from next-to-next-to-leading order in QCD perturbation theory.  We re-derived
this result in a manner that we find simple and transparent.

The gist of our approach  is the relation between
infrared-divergent contributions  to the manifestly finite  decay process $V \to q \bar q +X$   and
the production process $q \bar q \to V + X$ that can be studied using analytic continuation.
We have explicitly shown that while for the massless case this  analytic continuation is
harmless through NNLO, the situation is different in a massive theory.
There the phase from the analytic continuation of the one-loop soft
current combines with a non-trivial imaginary
part in the one-loop amplitude and gives   rise to an observable effect in  the cross-section.
Our derivation
provides a concrete and simple example of problems that one encounters when an analog of
a quantum mechanical Coulomb phase manifests itself
in massive non-abelian gauge theories. In fact,  it is relatively easy to
show that the offending phase  is related to a particular
double-particle massive cut that encapsulates the  long-distance interaction between
two incoming massive partons, see e.g.~\cite{Catani:2000ef}.

Before concluding, we briefly discuss the phenomenological implications of the above results.
One may argue that in collider phenomenology one does encounter processes
involving massive initial state quarks, e.g. $b\bar b\to H$ and similar.   In
fact, impressive machinery has been developed for dealing with such
processes~\cite{hfmatchings, recenthqworks}. However, in such  cases one always starts
with initial state gluons that subsequently split into a heavy
$b\bar b$ pair. It is important that massive quarks that originate in  such a splitting and  
participate in the hard scattering process after that are  {\it always} off-shell. Hence,  the average off-shellness
of initial state quarks that originate from  the gluon splitting  $g \to q \bar q$
provides a natural infrared cut-off for processes initiated by massive quarks.  For this reason, the infrared
divergence shown in Eq.~\eqref{eq3:16} can never appear in a realistic set up.

Nevertheless, computations with massive quarks in the initial state
can,  perhaps,  be used to test the sensitivity of partonic cross sections to infrared energy scales that the quark masses
may represent. For example, one may wonder to what extent the masses of the colliding quarks affect the
transverse momentum distributions of $Z$ and $W$ bosons at low $p_\perp$ -- a question, that may be quite relevant for
the determination of the $W$ boson mass at the LHC.  Our discussion
suggests  that, since  one starts being  sensitive to the off-shellness of quarks only
at $\mathcal O(m_q^4/m_V^4)$,  it should be possible
to develop a framework
 where one keeps track of terms of order $p_\perp/m_q\sim 1$ but neglects contributions  of order
$m_q^4/m_V^4$ and beyond.
 We leave this investigation, as well as the study of its potential phenomenological
applications, for the future.

\vspace*{1cm}
{\bf Acknowledgments}
DN would like to thank S.~Catani for useful discussions on the
original argument. We are grateful to S.~Forte, F.~Krauss, S.~Marzani and 
G.~Salam for many interesting
conversations. We would also like to thank Z.~Kunszt and S.~Marzani for a critical reading
of the manuscript. 
The research of FC is
partially supported by the ERC Starting Grant 804394 {\sc HipQCD}.
The research of KM is
partially supported by the Deutsche Forschungsgemeinschaft (DFG, German Research Foundation) under grant
396021762 - TRR 257.  DN is supported by the ERC Starting Grant REINVENT-714788.
LT is supported by the Royal Society through grant URF/R1/191125.

\appendix

\section{Analytic continuation of the one-loop integrals}
\label{sec:app}
In this appendix, we explicitly compute the analytic continuation of the
three  integrals $M_i$ given in Eq.~\eqref{eq:Mdef} under  the $p_{1,2}\to - p_{1,2}$ transformation.
We start with  decay kinematics, cf. Fig.~\ref{fig:soft}. Since
$s_{ij} = 2 p_i\cdot p_j$,  we find that under the
$p_{1,2}\to -p_{1,2}$ transformation, $s_{1g} \to -s_{1g}$, $s_{2g}\to-s_{2g}$ and $s_{12}\to s_{12}$. 
Therefore, to understand how the integrals change under analytic
continuation, we only need to study their dependence on $s_{1g},s_{2g}$. 

This is most easily achieved if we employ the Feynman-Schwinger parametrization for the integrals $M_{1,..,3}$. 
To derive a suitable representation, we start with the identity 
\begin{align}
\frac{1}{A_1 A_2 ... A_n} &= \Gamma(n) \prod_{i=1}^n \int_0^\infty  dx_i 
\frac{\delta(1-\sum_{j=1}^n x_j)}{\big[ \sum_{i=1}^n A_i x_i \big]^n} \nonumber \\
&= \Gamma(n) \left( \prod_{i \notin \Sigma}^{n} \int_0^\infty  dx_i \right)   \left(  \prod_{i \in \Sigma } \int_0^1  dx_i  \right) 
\frac{\delta(1-\sum_{j \in \Sigma}x_j)}{\big[ \sum_{i=1}^n A_i x_i \big]^n}\,,
\end{align}
where $\Sigma$ represents an arbitrary subset of $\left\{1,2,...,n\right\}$~\cite{ChengWu}.
For  each integrals $M_i$  we choose the set $\Sigma_i$ such that it contains the  Feynman
parameter  that is employed for the propagator $1/(-2p_2 \cdot k + i \delta)$, i.e. 
$\Sigma = \{3\}$ for $M_1$ and $M_2$ and $\Sigma=\{4\}$ for $M_3$.  We find 
\begin{align}
M_1(\ep,p_g;p_1,p_2) &=  - G_1(\epsilon) \prod_{i=1}^2 \int_0^\infty  dx_i \,  \frac{(x_1+x_2)^{-1+2\epsilon} }{
\big[  m^2 - s_{2g}\; x_2 - i \delta  \big]^{1+\epsilon} } \,, \nonumber \\
M_2(\ep,p_g;p_1,p_2) &=  - G_2(\epsilon) \prod_{i=1}^2 \int_0^\infty  dx_i \,  \frac{x_1^{-1+2\epsilon} }{
\big[  m^2 \lp1+x_2^2\rp - s_{1g}\; x_1 x_2   - s_{12}\; x_2 
 - i \delta \big]^{1+\epsilon}}\;
 \label{eq:A1}\\
M_3(\ep,p_g;p_1,p_2) &=  - G_3(\epsilon) \prod_{i=1}^3 \int_0^\infty  dx_i \; \frac{(x_1+x_2)^{2\epsilon} }{
\big[  m^2 \lp1+x_3^2\rp  - s_{1g}\; x_1 x_3  -s_{12}\; x_3 - s_{2g}\; x_2   - i \delta \big]^{2+\epsilon}}\;\,,
\nonumber
\end{align}
where the explicit form of the $G_i(\epsilon)$ is irrelevant in what follows.

It is straightforward to  study the dependence of the integrals on $s_{1g}$ and $s_{2g}$ using Eq.~\eqref{eq:A1}.
We begin with  $M_1$. By rescaling $x_i \to x_i/(-s_{2g}-i\delta)$ for $i=1,2$ we find
\begin{equation}
M_1(\ep,p_g;p_1,p_2) =   - G_1(\epsilon) \, (- s_{2g}-i\delta)^{-1-2 \epsilon} \prod_{i=1}^2 \int_0^\infty  dx_i  \; \frac{(x_1+x_2)^{-1+2\epsilon} }{
\left(  m^2 + x_2  \right)^{1+\epsilon}  } \,,
\label{eq:A2}
\end{equation}
so  that the entire dependence on $s_{2g}$  factorizes
\begin{equation}
M_1(\ep,p_g;p_1,p_2) \propto (- s_{2g}-i\delta)^{-1-2 \epsilon} = -|s_{2g}|^{-1-2\ep} e^{2i\pi\ep}.
\end{equation}
This implies 
\be
M_1(\ep,p_g;-p_1,-p_2) \propto  |s_{2g}|^{-1-2\ep},
\ee
and therefore
\be
M_1(\ep,p_g;-p_1,-p_2) = -M_1(\ep,p_g;p_1,p_2) e^{-2i\pi\ep}.
\label{eq:A5}
\ee
Furthermore, we note that Eq.~\eqref{eq:A2} implies that in the soft limit
\be
M_1(\ep,p_g;p_1,p_2) \sim E_g^{-2\ep}.
\label{eq:A6}
\ee

We analyse the integral  $M_2$ in a similar way.
In this case it is sufficient to rescale $x_1 \to x_1/(-s_{1g}-i\delta)$ to find
\begin{equation}
M_2(\ep,p_g;p_1,p_2) =   G_2(\epsilon) (  - s_{1g}-i\delta )^{-2 \epsilon } \prod_{i=1}^2 \int_0^\infty  dx_i \; \frac{x_1^{-1+2\epsilon} }{
\big[  m^2 \lp1+x_2^2\rp + x_1 x_2  - s_{12}\; x_2 -i\delta 
\big]^{1+\epsilon}}.\; 
\end{equation}
Hence, the dependence of $M_2$ on $p_1$ is governed by the following factor 
\begin{equation}
M_2(\ep,p_g;p_1,p_2) \propto (  - s_{1g}-i\delta )^{-2 \epsilon }. 
\label{eq:A8}
\end{equation}

Finally we discuss $M_3$. In this case, we rescale
$x_i \to x_i/(-s_{ig}-i\delta)$, where we stress that the rescaling is different for the two variables.
We obtain
\begin{align}
M_3(\ep,p_g;p_1,p_2) &=  - G_3(\epsilon) \prod_{i=1}^3 \int_0^\infty  dx_i \,
\frac{
\left( \frac{x_1}{-s_{1g}-i\delta}+\frac{x_2}{-s_{2g}-i\delta} \right)^{2\epsilon} 
}
{
\big[  m^2\lp1+x_3^2\rp   + x_2  -s_{12}\; x_3 + x_2x_3  -i\delta  \big]^{2+\epsilon}}.
\label{eq:A9}
\end{align}
Similarly to what was discussed for $M_1$, Eqs~(\ref{eq:A8}, \ref{eq:A9}) imply
\be
M_{2,3}(\ep,p_g;-p_1,-p_2) = M_{2,3}(\ep,p_g;p_1,p_2)e^{-2i\pi\ep},
\label{eq:A10}
\ee
and
\be
M_{2,3}(\ep,p_g;p_1,p_2) \sim E_g^{-2\ep}.
\label{eq:A11}
\ee

Finally, we note that 
Eq.~\eqref{eq:gMdef} along with Eqs~(\ref{eq:fdef}, \ref{eq:A5}, \ref{eq:A6}, \ref{eq:A10}, \ref{eq:A11})
imply that
\be
g_{12}^{(1)}(\ep,p_g;-p_1,-p_2) = e^{-2i\ep\pi} g_{12}^{(1)}(\ep,p_g;p_1,p_2),
\ee
and
\be
g_{12}^{(1)}(\ep,p_g;p_1,p_2) = \frac{\alpha_s}{2\pi}
E_g^{-2\ep}
\sum_{k=-2}^{\infty} \left[
\mathfrak r_{k} + i\cdot \mathfrak i_{k}
\right]\ep^k,
\ee
with $\mathfrak r_{k}$ and $\mathfrak i_k$ analytic in $E_g$.  These formulas are used in the main body of the paper
to explain the appearance of non-cancelling infrared divergencies in collisions of two massive quarks.

\end{document}